\documentstyle[twoside,fleqn,espcrc2,psfig]{article}
\def\be{\begin{eqnarray}}
\def\ee{\end{eqnarray}}
\def\ep{\varepsilon}

\newcommand{\psla}{p\hspace{-0.45em}/}

\title{Electroweak corrections to the muon anomalous magnetic moment}
\author{Andrzej Czarnecki and Bernd Krause
\address{Institut f\"ur Theoretische Teilchenphysik, \\
Universit\"at Karlsruhe,
D-76128 Karlsruhe, Germany\\
Report-no: TTP96-21}}

\begin{document}

\begin{abstract}
The two-loop electroweak radiative corrections to the muon's
anomalous magnetic moment, $a_\mu\equiv (g_\mu-2)/2$, are presented.
We obtain an overall 22.6\%
reduction in the 
electroweak contribution $a_\mu^{\rm EW}$ from $195\times 10^{-11}$ to 
$151(4)\times 10^{-11}$.  Implications for the full standard model
prediction and
an upcoming high precision measurement of $a_\mu$ are briefly discussed.
Some aspects of the calculations are discussed in detail.
\end{abstract}

\maketitle

The anomalous magnetic moment of the muon, $a_\mu\equiv (g_\mu-2)/2$,
provides both a sensitive quantum loop test of the standard
$SU(3)_C\times SU(2)_L\times U(1)$ model and a window to potential
``new physics'' effects.  The current experimental average \cite{PDG}
\begin{eqnarray}
a_\mu^{\rm exp} = 116592300 (840) \times 10^{-11}
\end{eqnarray}
is in good agreement with theoretical expectations and already
constrains physics beyond the standard model such as supersymmetry and
supergravity
\cite{km90,Nath95}, dynamical or loop muon mass generation \cite{3},
compositeness \cite{4}, leptoquarks \cite{Couture95} etc.

An upcoming experiment E821
\cite{Hughes92} at Brookhaven National Laboratory is
expected to start taking data in 1997.  With one month
of dedicated running, it is expected to reduce the uncertainty 
in $a_\mu^{\rm exp}$ to roughly  $\pm 40\times 10^{-11}$, 
more than a factor of 20 improvement.
With subsequent longer dedicated runs it could statistically 
approach the anticipated
systematic uncertainty of about $\pm 10-20\times 10^{-11}$ \cite{Bunce}.
At those levels, both electroweak
one and two loop effects become important and ``new physics'' at the
multi-TeV scale is probed.  Indeed, generic muon mass generating
mechanisms (via perturbative or dynamical loops \cite{3}) lead to
$\Delta a_\mu \approx m_\mu^2/\Lambda^2$, where $\Lambda$ is the scale
of ``new physics''.  At $\pm 40\times 10^{-11}$ sensitivity, $\Lambda
\approx 5$ TeV is being explored.

To fully exploit the anticipated experimental improvement, the
standard model prediction for $a_\mu$ must be known with comparable
precision.  That requires detailed studies of very high order QED
loops, hadronic effects, and electroweak contributions through two
loop order.  The contributions to $a_\mu$ are traditionally divided
into 
\begin{eqnarray}
a_\mu=a_\mu^{\rm QED}+a_\mu^{\rm Hadronic}+a_\mu^{\rm EW}
\end{eqnarray}
QED loops have been computed to very high order
\cite{r4,Kinoshita95,RemRhein}
\begin{eqnarray}
a_\mu^{\rm QED} &=& 116584706(2)\times 10^{-11}.
\end{eqnarray}
The  uncertainty of the QED result 
is well within the $\pm 20-40\times 10^{-11}$ goal.

Hadronic vacuum polarization corrections to $a_\mu$ enter at ${\cal
O}(\alpha/\pi)^2$.  They can be evaluated via a dispersion relation
using $e^+e^- \to hadrons$ data and perturbative QCD (for the
very high energy regime).  Employing a recent analysis of $e^+e^-$
data \cite{Jeg95,JegRhein} along with an estimate of the leading ${\cal
O}(\alpha/\pi)^3$ effects, we find  \cite{CKM95}
\begin{eqnarray}
a_\mu^{\rm Hadronic}({\rm vac.\,pol.})= 6934(153)\times 10^{-11}
\label{eq:hadr}
\end{eqnarray}
Unfortunately, the error has not yet reached the desired level of
precision.  Ongoing improvements in $e^+e^- \to hadrons$ 
measurements at low energies along with additional theoretical input
should significantly lower the uncertainty
in (\ref{eq:hadr}). Nevertheless, reducing the hadronic error below
$\pm 20 \times 10^{-11}$ or even $\pm 40 \times 10^{-11}$ 
remains a formidable challenge.

The result in (\ref{eq:hadr}) must be supplemented by hadronic light
by light amplitudes (which are of three loop origin)
\cite{light,haya95,Bijnens95}.  Here, we employ
a recently updated 
study by Hayakawa, Kinoshita, and Sanda \cite{haya95} which
gives
\begin{eqnarray}
a_\mu^{\rm Hadronic}({\rm light\, by \, light})= -52(18)\times
10^{-11}
\label{eq:light}
\end{eqnarray}
However, we note that the result is somewhat dependent on the low
energy model of hadronic physics employed and  continues to be
scrutinized.  Combining (\ref{eq:hadr}) and (\ref{eq:light})
leads to the total hadronic contribution
\begin{eqnarray}
a_\mu^{\rm Hadronic}= 6882(154)\times 10^{-11}
\end{eqnarray}

Now we come to the electroweak contributions to $a_\mu$, the main
focus of our work and the impetus for forthcoming experimental effort.
At the one loop level, the standard model predicts 
\cite{fls72,Jackiw72,ACM72,Bars72,Bardeen72}
\begin{eqnarray}
\lefteqn{a_\mu^{\rm EW}(\rm 1\,loop) =
{5\over 3}{G_\mu m_\mu^2\over 8\sqrt{2}\pi^2}}
\nonumber\\ && \times
\left[1+{1\over 5}(1-4s_W^2)^2
+ {\cal O}\left({m_\mu^2 \over M^2}\right) \right]
\nonumber \\
&& \approx 195 \times 10^{-11}
\label{eq:oneloop}
\end{eqnarray}
where $G_\mu = 1.16639(1) \times 10^{-5}$ GeV$^{-2}$, $M=M_W$ or
$M_{\rm Higgs}$, and the weak mixing angle
$\sin^2\theta_W\equiv s_W^2 = 1-M_W^2/M_Z^2=0.224$.  We can safely
neglect the ${\cal O}\left({m_\mu^2 / M^2}\right)$ terms in
(\ref{eq:oneloop}).

\begin{figure}
\begin{minipage}{16.cm}
\hspace*{.8cm}
\[
\mbox{
\begin{tabular}{ccc}
\psfig{figure=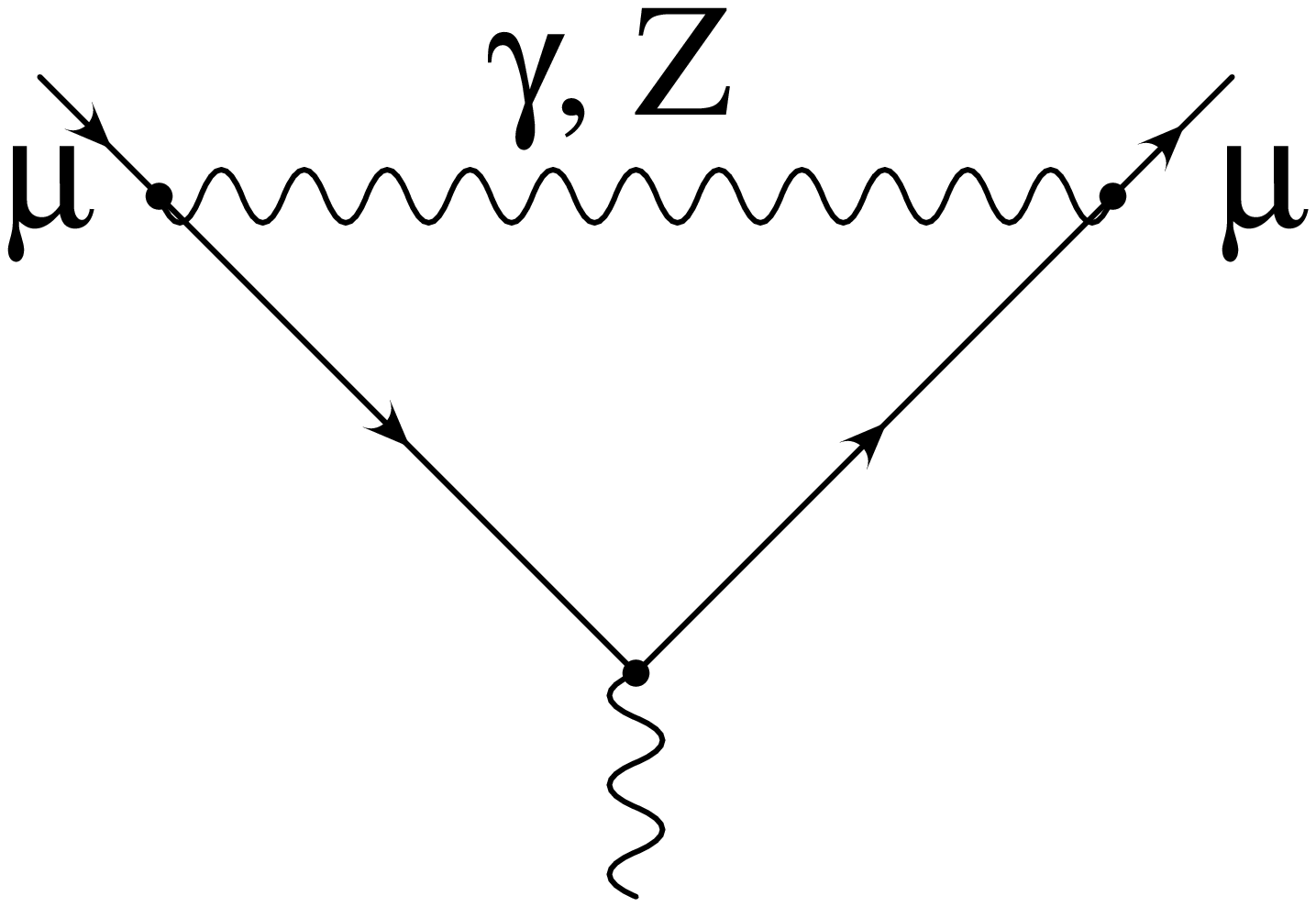,width=20mm,bbllx=210pt,bblly=410pt,bburx=630pt,bbury=550pt} 
&\hspace*{.2cm}
\psfig{figure=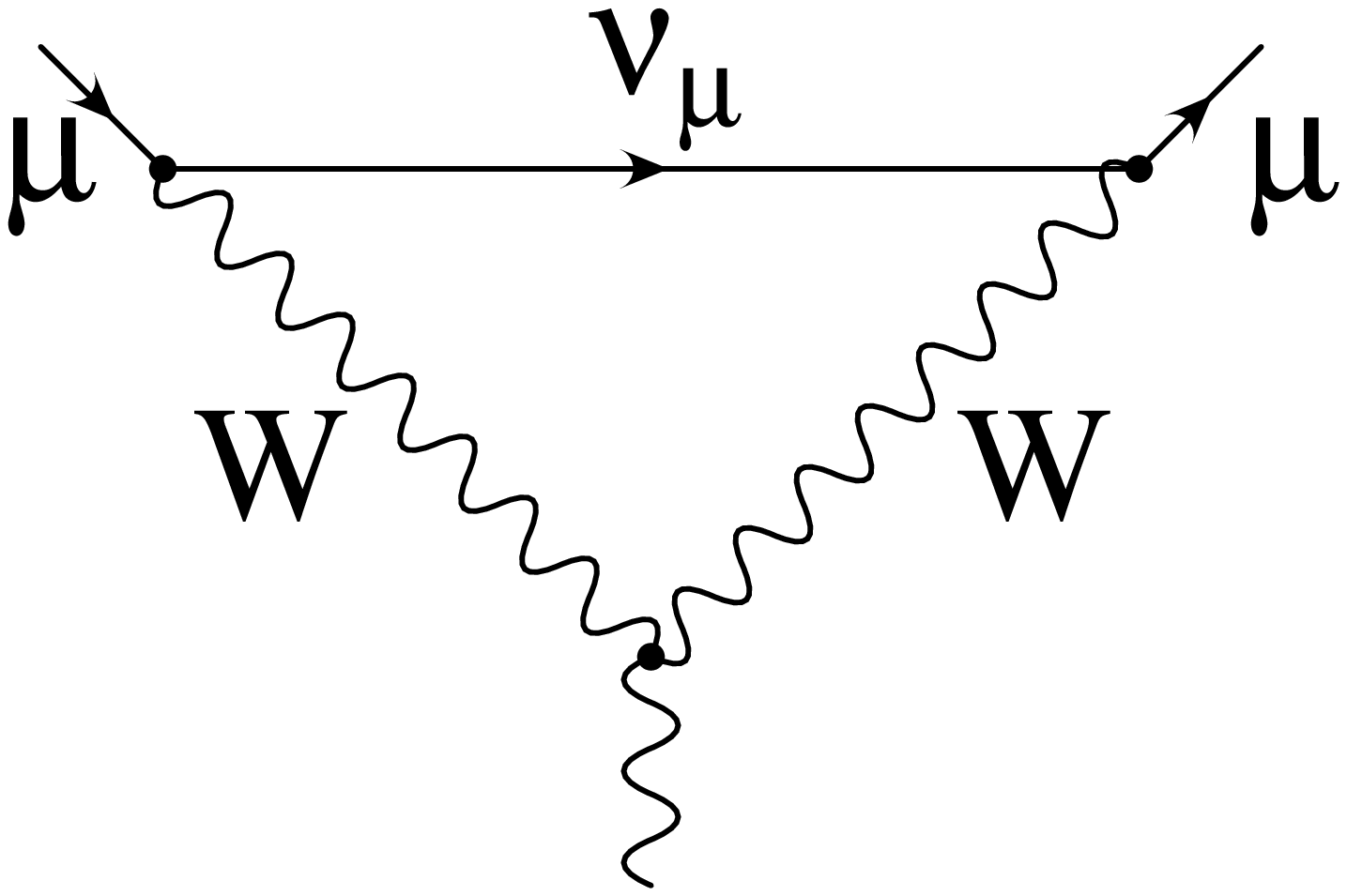,width=20mm,bbllx=210pt,bblly=410pt,bburx=630pt,bbury=550pt}
&\hspace*{.2cm}
\psfig{figure=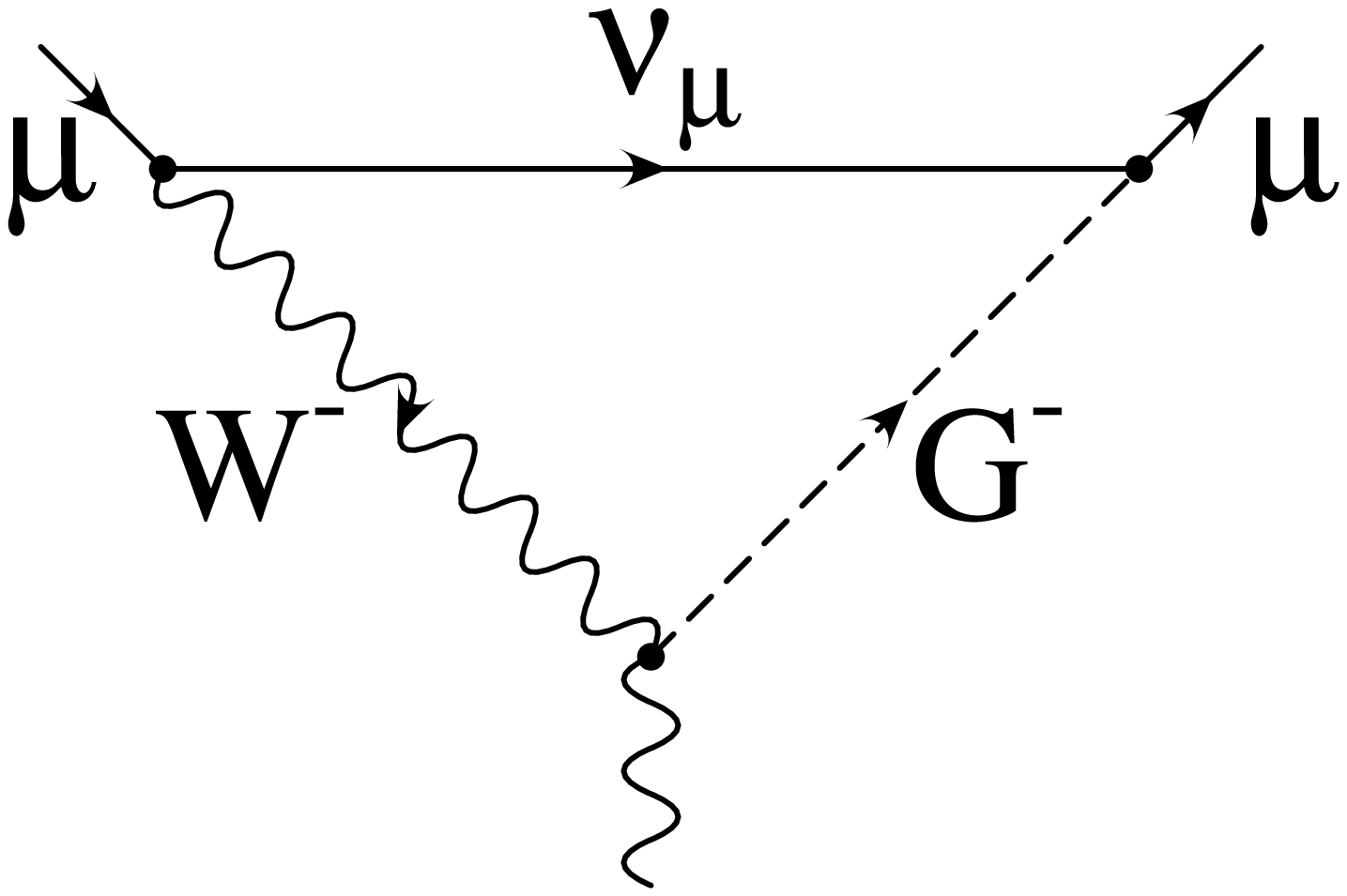,width=20mm,bbllx=210pt,bblly=410pt,bburx=630pt,bbury=550pt}
\\
[3mm]
&&\\
\psfig{figure=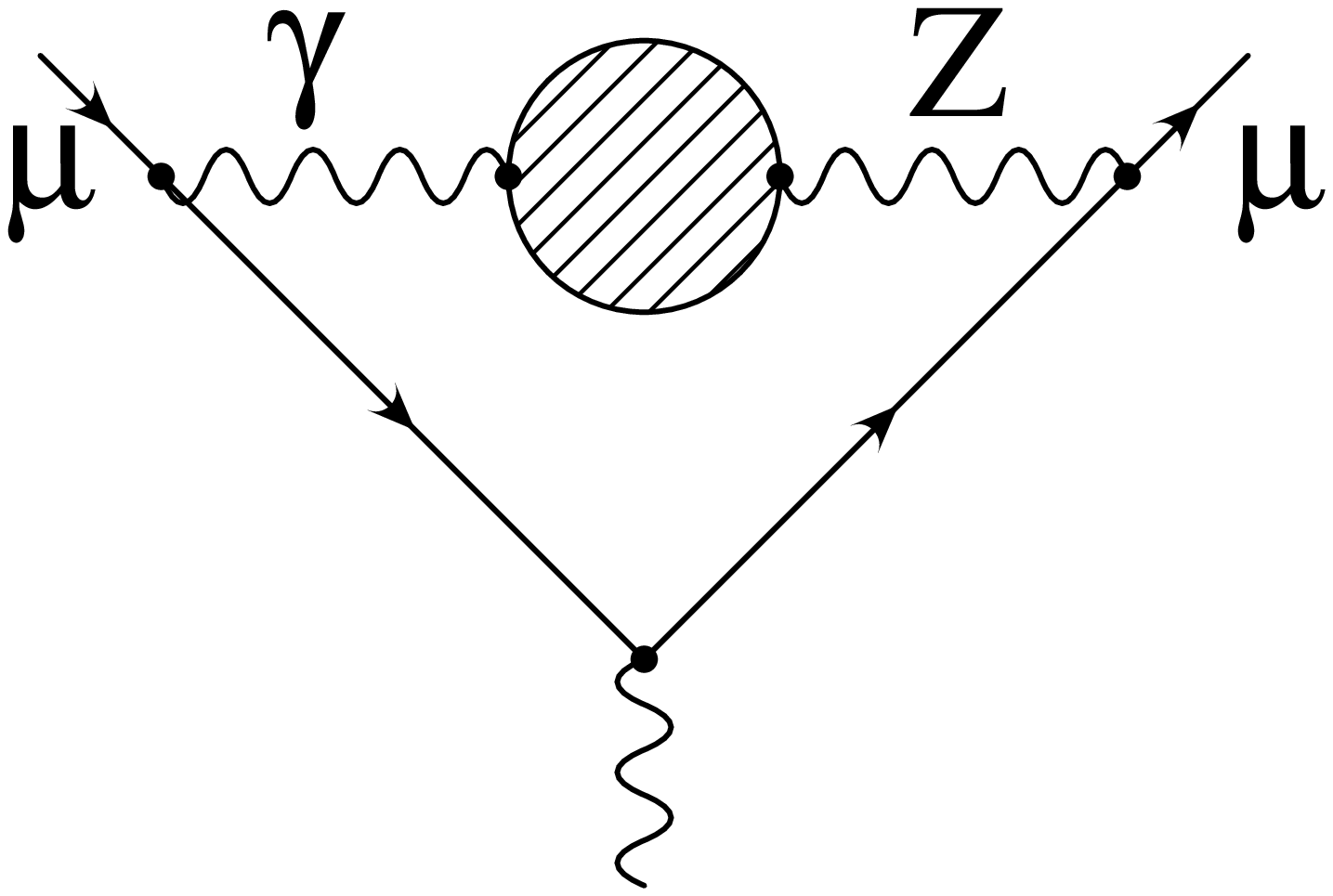,width=20mm,bbllx=210pt,bblly=410pt,bburx=630pt,bbury=550pt} 
&\hspace*{.2cm}
\psfig{figure=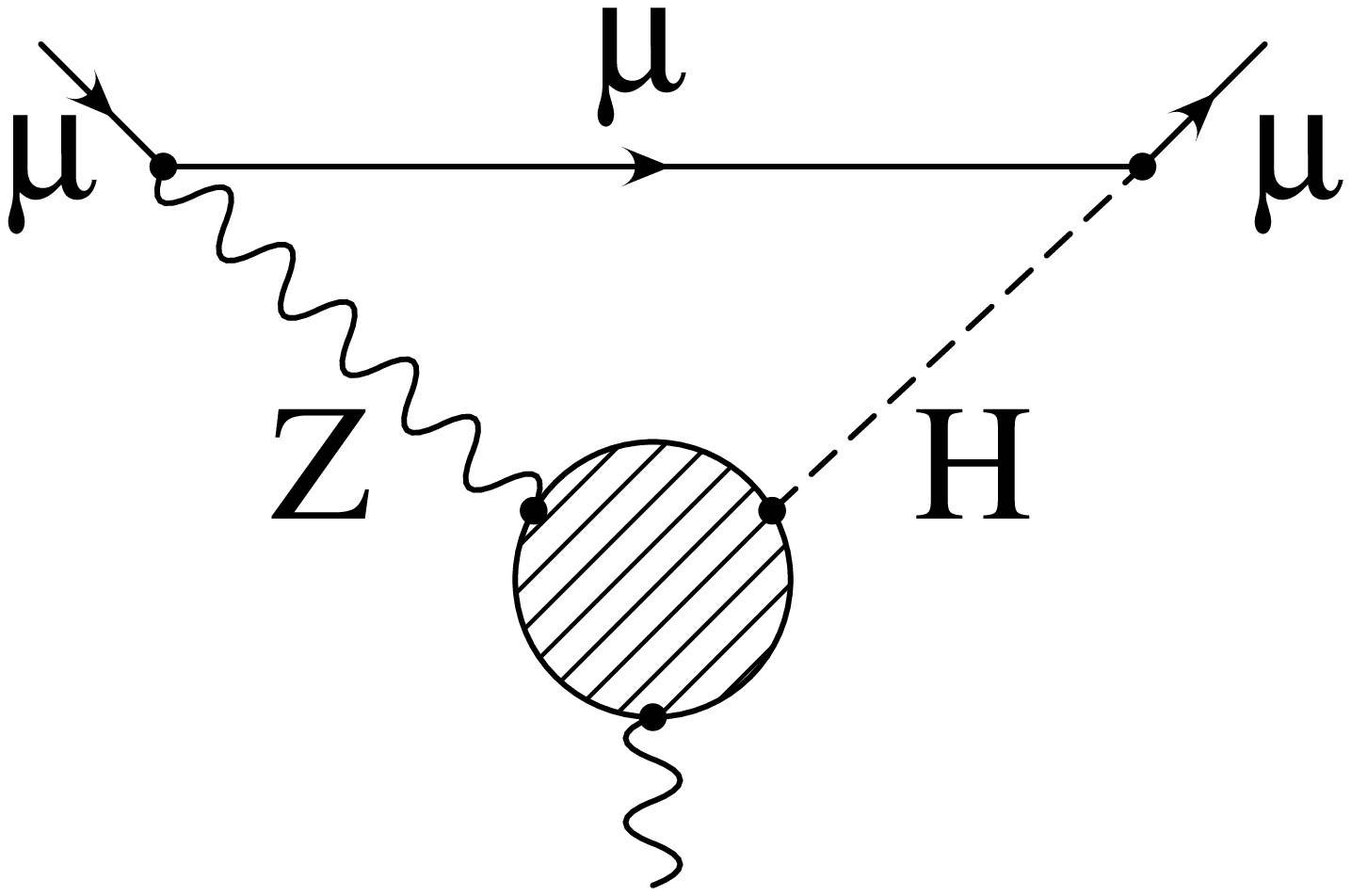,width=20mm,bbllx=210pt,bblly=410pt,bburx=630pt,bbury=550pt}
&\hspace*{.2cm}
\psfig{figure=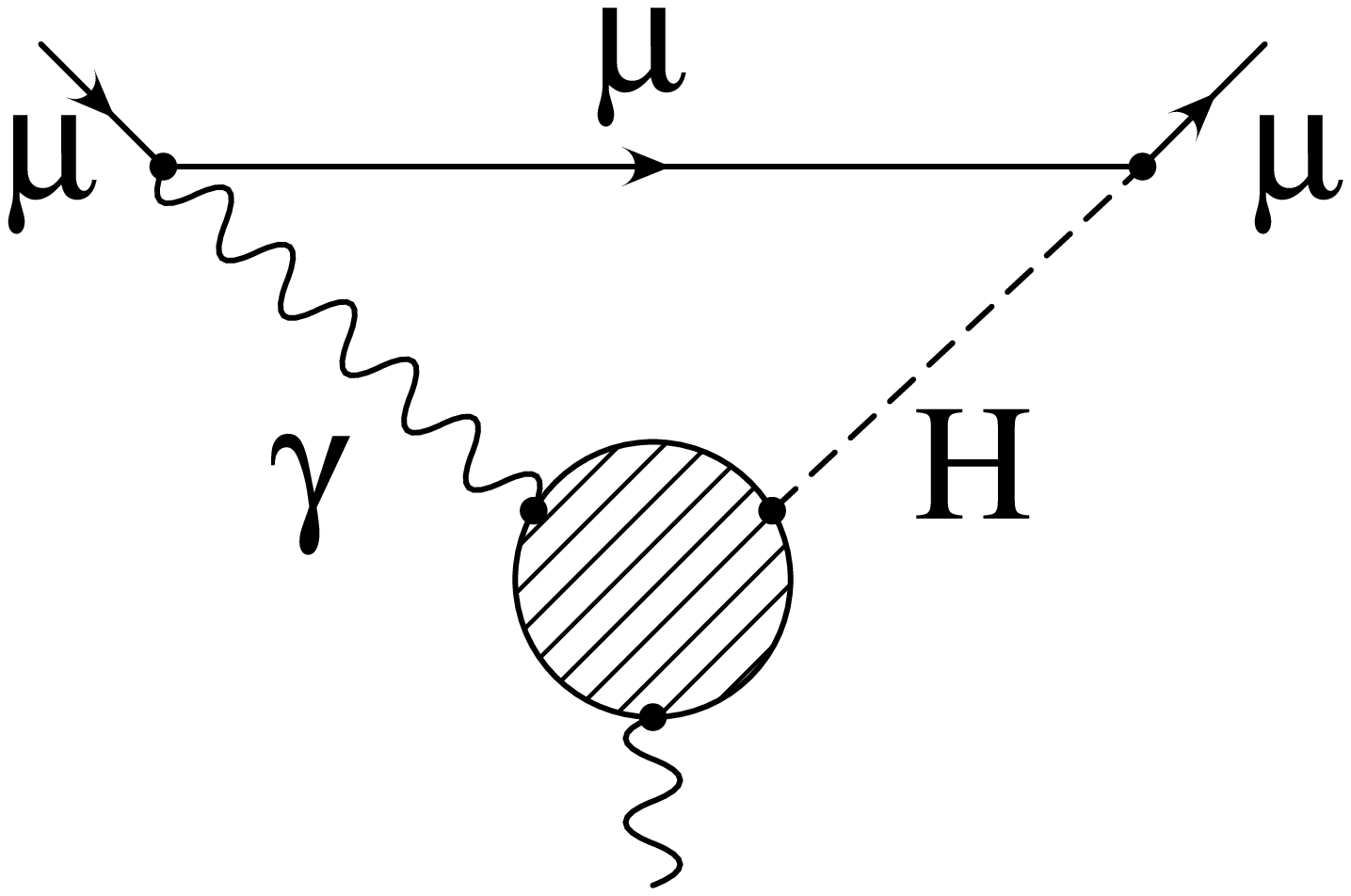,width=20mm,bbllx=210pt,bblly=410pt,bburx=630pt,bbury=550pt}
\\
[3mm]
&&\\
\psfig{figure=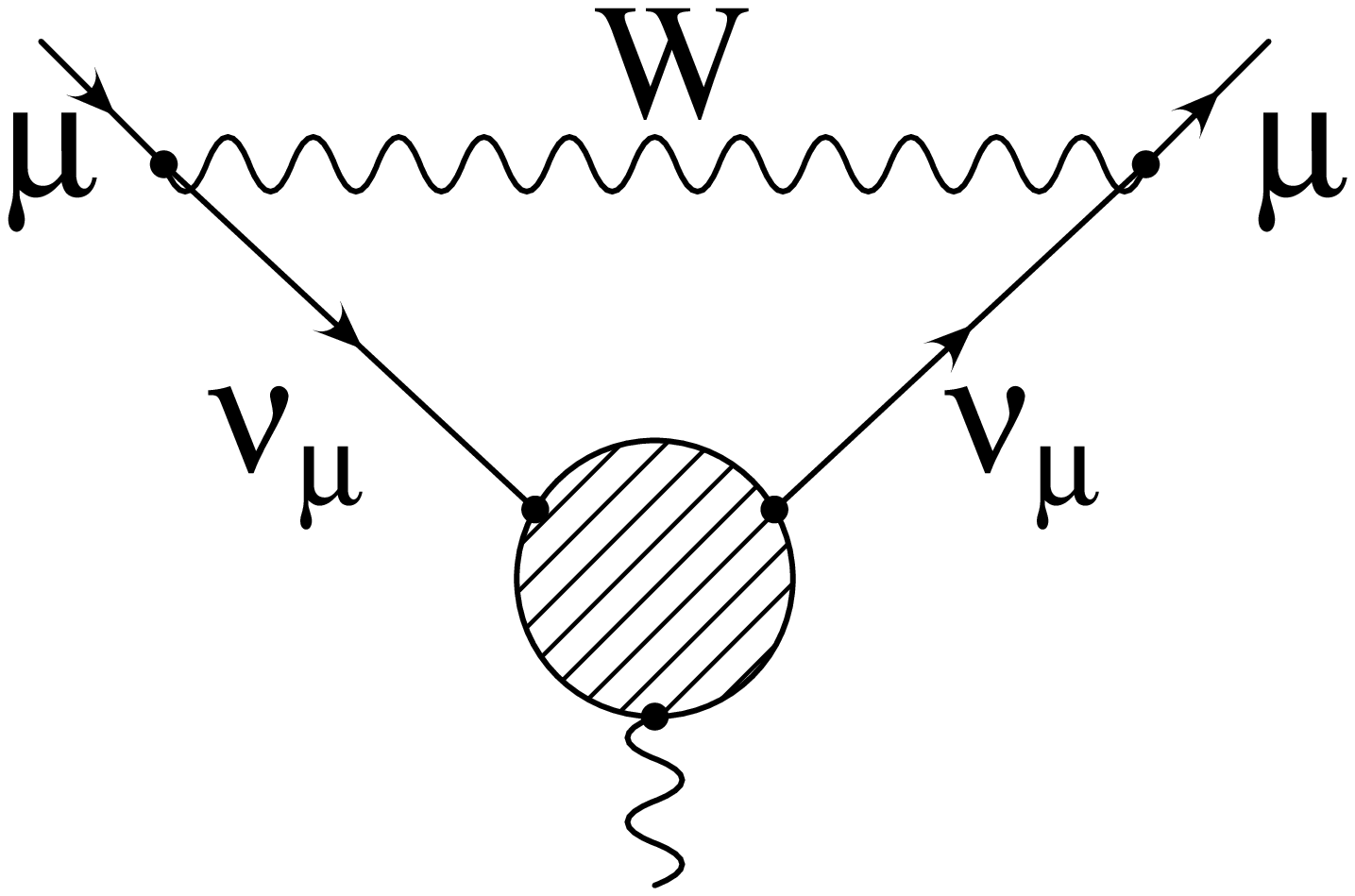,width=20mm,bbllx=210pt,bblly=410pt,bburx=630pt,bbury=550pt} 
&\hspace*{.2cm}
\psfig{figure=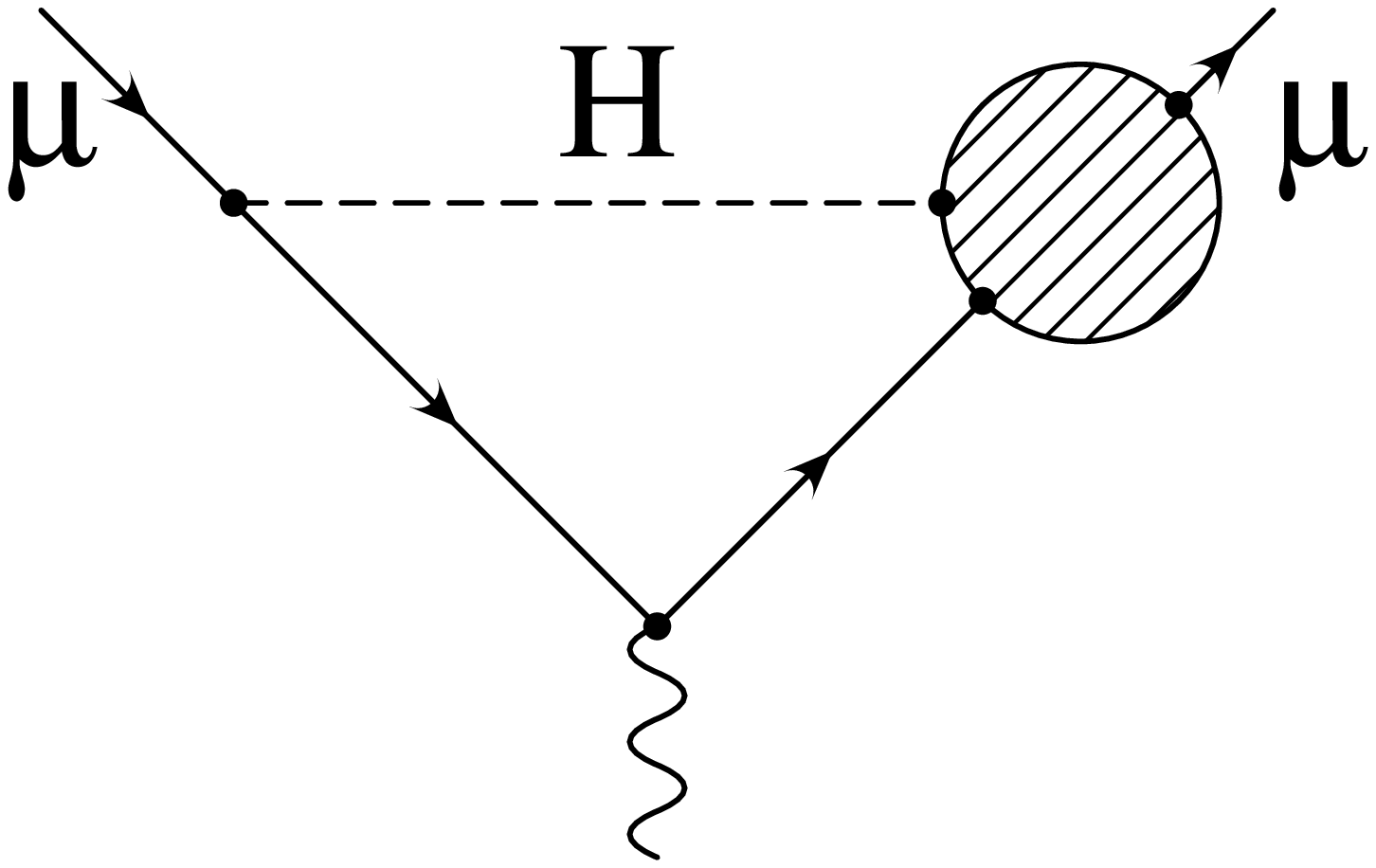,width=20mm,bbllx=210pt,bblly=410pt,bburx=630pt,bbury=550pt}
&\hspace*{.2cm}
\psfig{figure=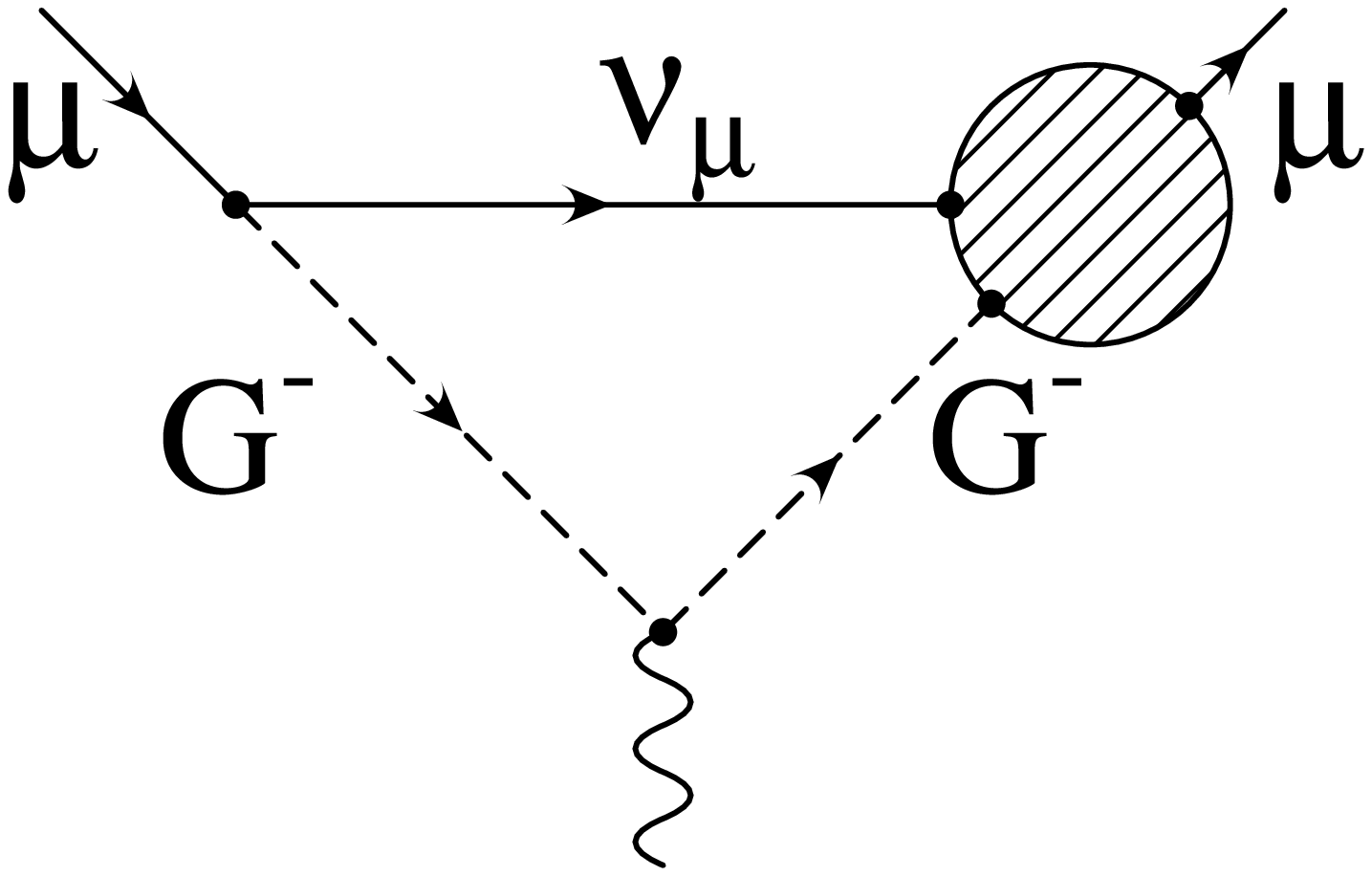,width=20mm,bbllx=210pt,bblly=410pt,bburx=630pt,bbury=550pt}
\\
[3mm]
&&\\ 
\end{tabular}
}
\]
\end{minipage}
\caption{\label{bosonen}Bosonic diagrams contributing to g-2 of the muon }
\end{figure}

The one loop result in (\ref{eq:oneloop}) is about five to ten times the
anticipated experimental error.  Naively, one might expect higher order (2
loop) electroweak contributions to be of relative ${\cal O}(\alpha/
\pi)$ and hence negligible; however, that is not the case.  Kukhto,
Kuraev, Schiller, and Silagadze (KKSS)  \cite{KKSS} have shown that
some two loop electroweak contributions can be quite large and must be
included in any serious theoretical estimate of $a_\mu^{\rm EW}$ or
future confrontation with experiment.  Given the KKSS observation, a
detailed evaluation of the two loop electroweak contributions to
$a_\mu$ is clearly warranted.  We have reported  the complete results of
such an analysis in two recent papers \cite{CKM95,CKM96}.
The types of diagrams contributing to two-loop electroweak corrections
are shown in \ref{bosonen}. These fall into a number of different categories.
Diagrams with a closed fermion loop,
together with a class of bosonic diagrams, represent
corrections to vertices and propagators of the one-loop 
electroweak diagrams (shown in the first line in Fig.\ref{bosonen}). 

\begin{figure}
\begin{minipage}{16.cm}
\hspace*{.8cm}
\begin{tabular}{cc}
\psfig{figure=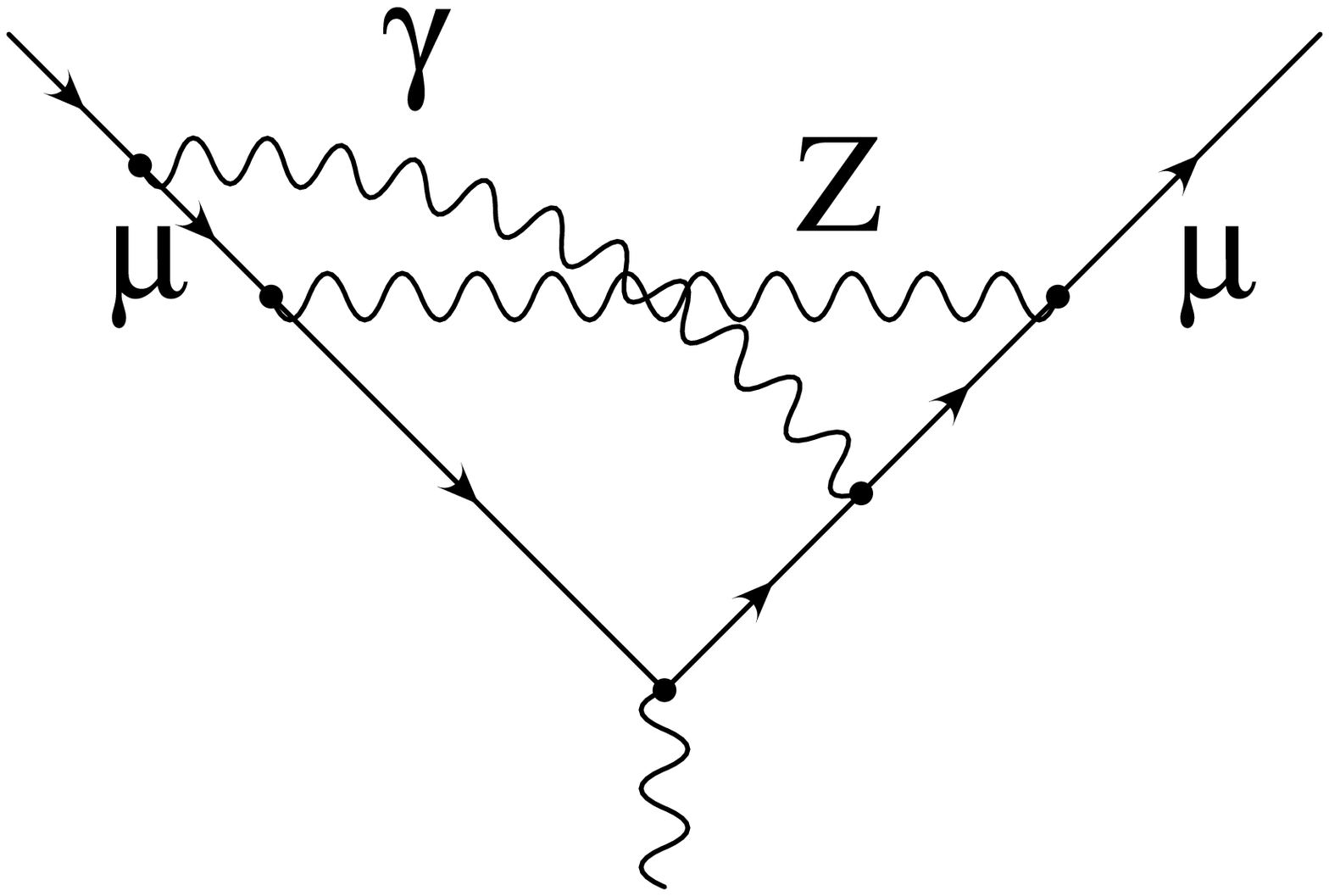,width=20mm,bbllx=210pt,bblly=410pt,bburx=630pt,bbury=550pt} 
&\hspace*{.2cm}
\psfig{figure=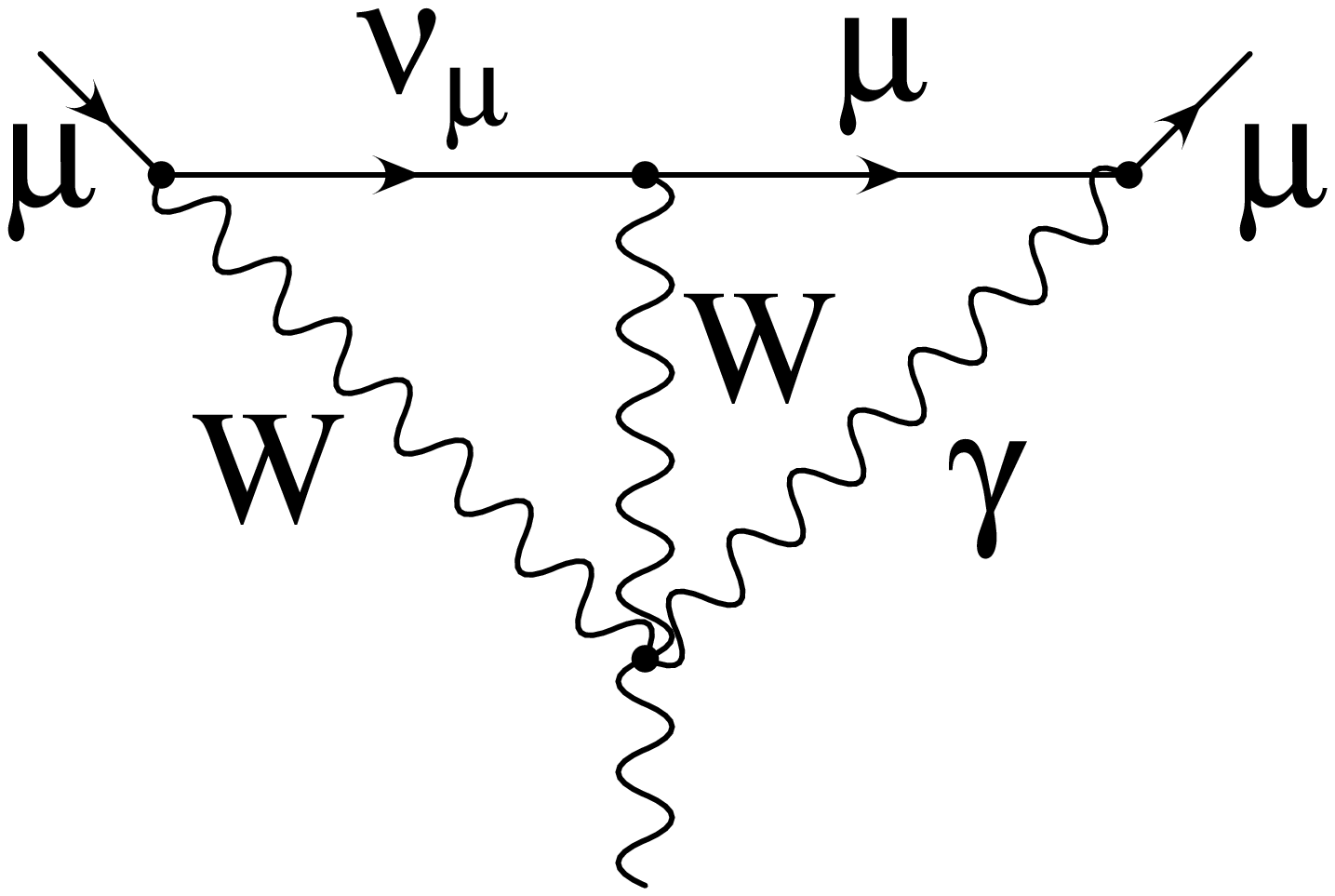,width=20mm,bbllx=210pt,bblly=410pt,bburx=630pt,bbury=550pt}
\\
\end{tabular}
\end{minipage}
\caption{\label{crossed}Examples of non-planar diagrams and diagrams
with quartic couplings.}
\end{figure}

In addition we have to calculate new types of diagrams which appear at
the two-loop level, such as $\gamma Z$ mixing, an induced $\gamma \gamma H$
vertex etc. (lines two and three of Fig.\ref{bosonen}).
Finally, there are non-planar diagrams and diagrams with quartic
couplings; some examples are shown in Fig.\ref{crossed}.

\vspace{.5cm}

The final result of the total electroweak contributions to the g-2 of
leptons -- including one- and two-loop corrections -- is
\be
a_e^{\rm EW}&=&3.0\cdot 10^{-14}\nonumber\\
a_\mu^{\rm EW}&=&151\cdot 10^{-11}\\
a_\tau^{\rm EW}&=&4.7\cdot 10^{-7} \; .\nonumber
\ee
In this paper we discuss some details of the procedures we have
employed in that project.  In particular we discuss the method for
projecting out the Pauli formfactor in $d$ dimensions, and some
aspects of the fermionic contributions.

\subsection*{Diagrams with charged bosons}

Contributions of diagrams with a fermion loop connected to the muon
line via two charged bosons are shown in Fig. 3.
Fermions with isospin $\pm 1/2$ are denoted by $u$ and $d$,
respectively. Of course, in the case of leptons diagrams (a) and (f)
vanish. 

We first consider the case when the fermions in the loop belong to the
first two generations. Here the masses of the fermions in the loop do
not influence the result very much and we neglect them. The ratio of
the neglected terms to the result is at most of the order of
\be
{m_c^2\over M_W^2}\ln{ M_W^2\over m_c^2}<0.3\%
\ee
with $m_c$ denoting the mass of the charm quark $\approx 1.5 {\rm
GeV}$. In the massless approximation we need to consider only the
first four diagrams in Fig. 3. In ref.\cite{KKS} it has been argued that the
diagrams 1(a) and 1(b) vanish by virtue of Furry's theorem. We find
that this is not true even after adding contributions of all fermions
in a generation. Furry's theorem consists in the observation that
the sum of contributions of diagrams with two different orientations of the
fermion loop vanishes. This is not the case for diagrams 1(a)
and 1(b) because for every fermion flavor interacting with the
external photon there is only one possible orientation of the fermion
line. Only those parts of the expressions which contain a single
$\gamma_5$ cancel out after adding contributions from the up-type
quark, down-type quark and from the lepton.

Let us consider diagrams 1(a) and 1(b) in more detail.
The contributions of these diagrams can be
divided up into those with an odd number of vector couplings to the fermion
loop ($VVV$ or $VAA$) and with two vector, one axial vector
coupling ($AVV$). The dominant contribution is obtained
by taking massless fermions in the loop. For
diagrams of type 1(a) and 1(b) we find (the subscript indicates the
fermion to which the photon couples)

\be
\Delta a_{f}^{VVV}  &=& \!\!\! {C N_C^f\over s_W^2} Q_f I_f\left({19\over 36}+{7\over 18\ep}-{7\over 9}\ln M_W^2 \right) \nonumber \\
\Delta a_{f}^{AVV}&=& {C N_C^f \over s_W^2} Q_f\left(+{1\over 4}\right)
\ee
with
\be
C = {G_\mu \alpha m_\mu^2\over 8\sqrt{2}\pi^3} \; .
\ee
Here we used the shorthand notation $VVV$ for the sum of $VVV$+$VAA$. 
$N_C^{quarks}=3$ 
is the number of colors (with $N_C^{lepton}=1$ understood) and 
$Q_f$ denotes the fermion charge.
Only the parity conserving part contains divergences; they are
proportional to the lowest order 
$\gamma W W$ vertex.

Summing the $AVV$ part over one light generation (l.g.) we obtain 
\be
\Delta a_{l.g.}^{AVV} = 0
\ee
due to the fact that in the standard model
\be
\sum_f N^f_C Q_f=0 \; .
\ee
The $VVV$ contribution of one light generation is 
\be
\Delta a_{l.g.}^{VVV}&=&
 {C \over s_W^2}
\sum_f N^f_C Q_f I_f \times \\
&& \times\left({19\over 36}+{7\over 18\ep}-{7\over 9}\ln M_W^2 \right)
\nonumber \\
&=&{C \over s_W^2}\cdot 2\cdot
 \left({19\over 36}+{7\over 18\ep}-{7\over 9}\ln M_W^2 \right)\nonumber
\ee

This example demonstrates explicitly that the contributions from three
vector couplings to the fermion loop do not cancel. We would like to
emphasize that this has nothing to do with the masses inside the fermion
loop. We can put them equal (even equal to zero as in our example)
just like in the QED case.
The Furry theorem is not applicable because, unlike in QED, the charge flow
through the fermion loop destroys the symmetry between the fermion
loop diagram and the diagram with reversed fermion direction.

We summarize now the total contributions from a single light generation
(where appropriate, we always include a mirror counterpart 
of the diagrams)

\be
\Delta a_{1a}&=& {C\over s_W^2} Q_u\left({37\over 72}+{7\over 36\ep}
-{7\over 18}\ln M_W^2 \right) \nonumber \\
\Delta a_{1b}&=& {C\over s_W^2} Q_d\left(-{1\over 72}-{7\over 36\ep}
+{7\over 18}\ln M_W^2 \right) \nonumber \\
\Delta a_{1c}&=& {C\over s_W^2} \left({1\over 54}-{1\over 6\ep}
+{1\over 3}\ln M_W^2 \right) \nonumber \\
\Delta a_{1d}&=& {C\over s_W^2} \left(-{1\over 216}-{1\over 36\ep}
+{1\over 18}\ln M_W^2 \right) 
\label{light}
\ee

These expressions have to be multiplied by a color factor where necessary.
We put the relevant Kobayashi-Maskawa
matrix elements equal to 1. Adding contributions of all fermions we
obtain for a single light generation
\be
\Delta a_{light}&=& {C\over s_W^2} {10\over 9}
\ee

We now proceed to the contribution of the third generation. For the
$\tau$ lepton loop we can still use eq. (\ref{light}). We obtain
\be
\Delta a_{\tau}&=& {C\over s_W^2} {1\over 36}
\ee

In diagrams with quarks we neglect the mass of the $b$. To leading
order in ${M_W^2\over m_t^2}$ we obtain for the sum of all diagrams
containing top and bottom quark loops
\be
\lefteqn{\Delta a_{tb} =  {C N_C\over s_W^2} \left[
 -{1\over 36}-{1\over 6}\ln {m_t^2\over M_W^2} 
 \right.} \\
&&\hspace{.5cm}\left. +{m_t^2\over M_W^2}\left(-{8\over 9}
-{5\over 12\ep}+{5\over 6}\ln m_t M_W\right)\right]\nonumber
\ee

The singular terms ${m_t^2\over \ep} $ are cancelled by
renormalization of the W boson mass present in the one loop
electroweak contributions to muon g-2.

\vspace{1cm}
%
\noindent
\begin{minipage}{10.cm}
\[
\mbox{
\hspace*{1cm}
\begin{tabular}{cc}
\psfig{figure=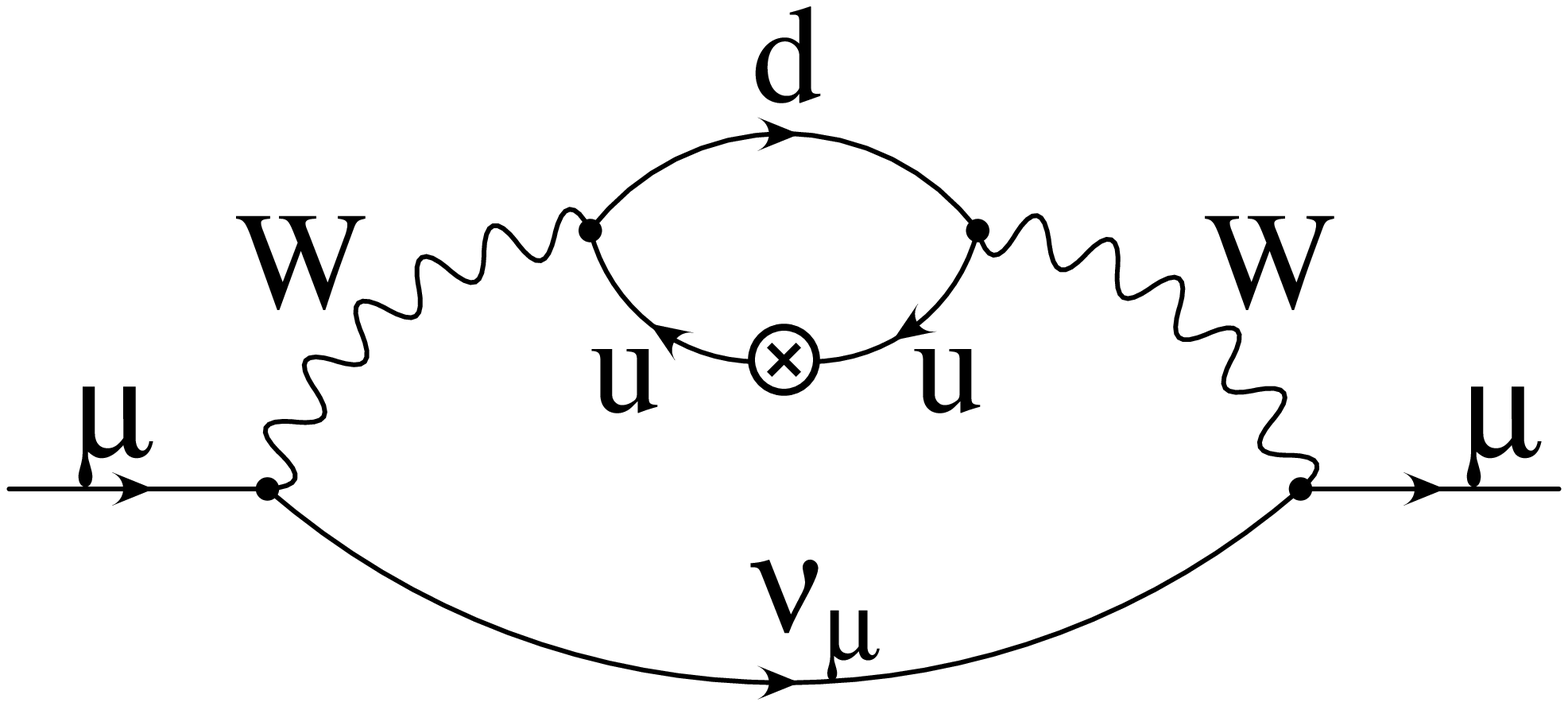,width=20mm,bbllx=210pt,bblly=410pt,bburx=630pt,bbury=550pt} 
&\hspace*{.6cm}
\psfig{figure=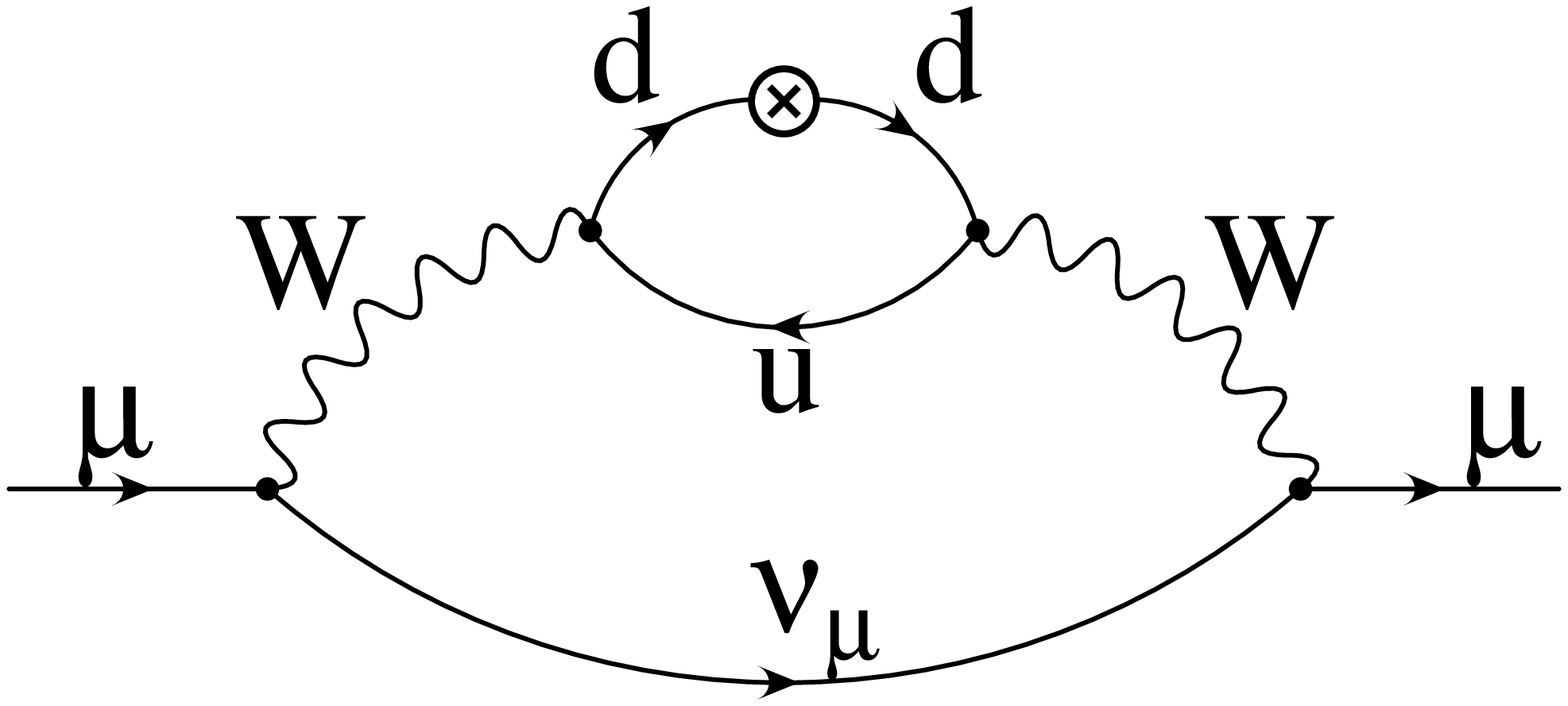,width=20mm,bbllx=210pt,bblly=410pt,bburx=630pt,bbury=550pt}
\\[5mm]
\rule{-13mm}{0mm} (a) &\hspace*{-.1cm} \rule{-4mm}{0mm}(b) 
\end{tabular}}
\]
\end{minipage}
\vspace*{1mm}

\noindent
\begin{minipage}{16.cm}
\[
\mbox{
\hspace*{1cm}
\begin{tabular}{cc}
\psfig{figure=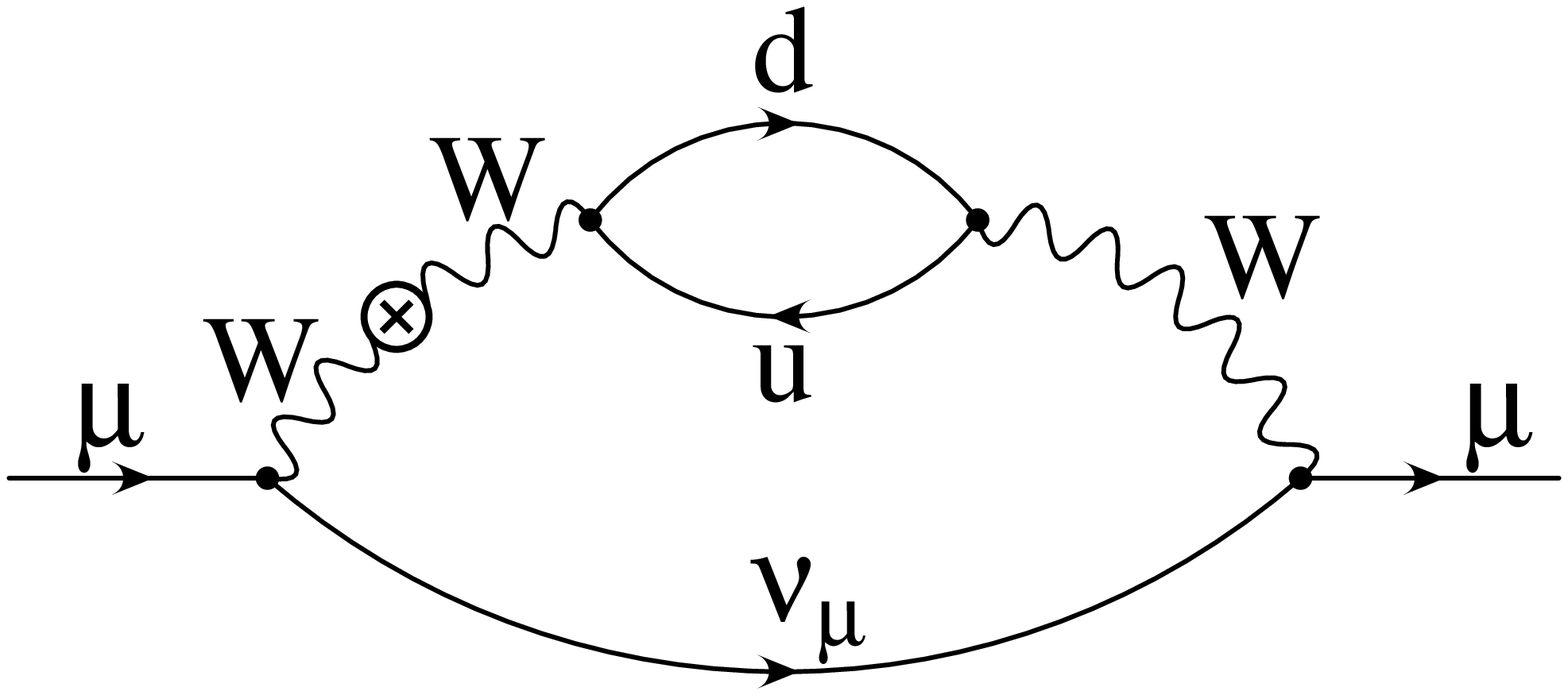,width=20mm,bbllx=210pt,bblly=410pt,bburx=630pt,bbury=550pt}
&\hspace*{.6cm}
\psfig{figure=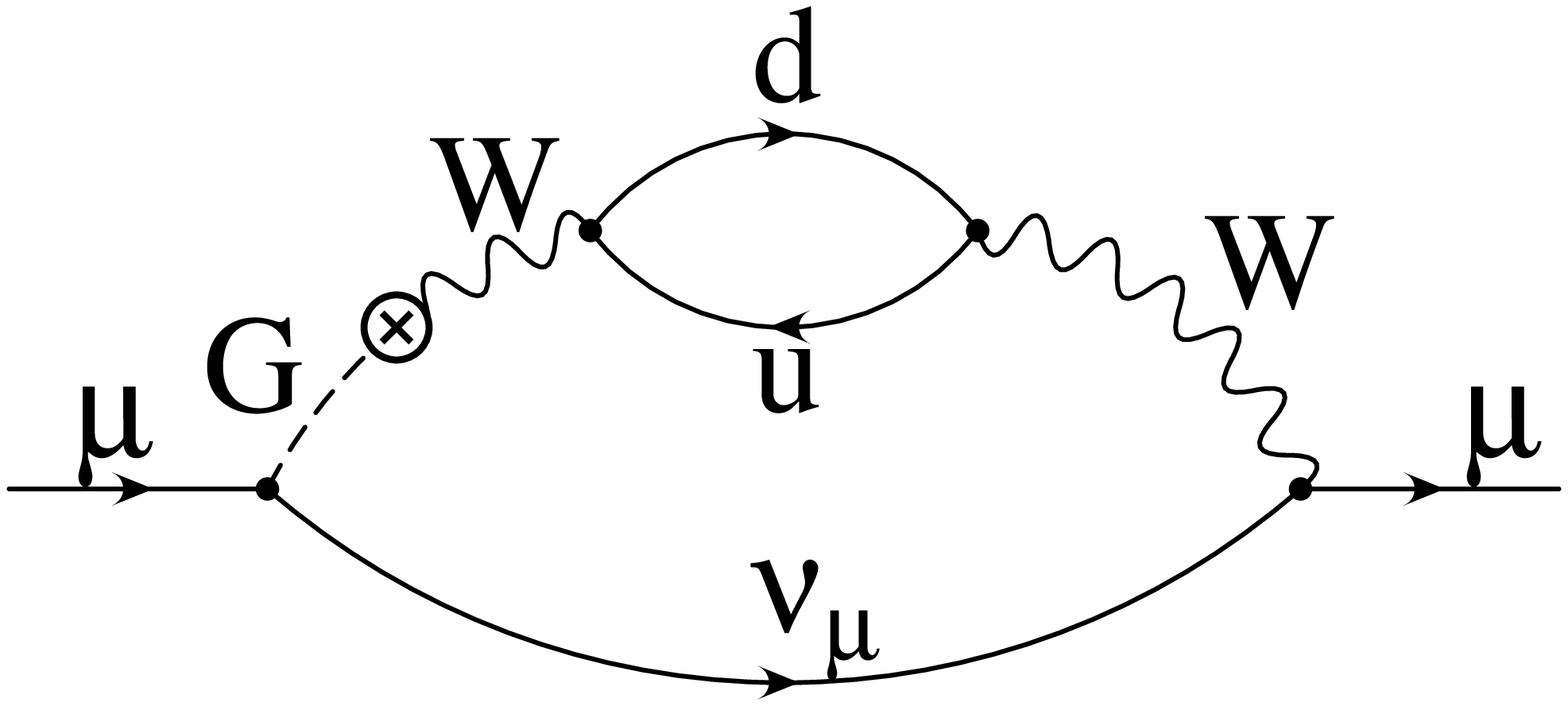,width=20mm,bbllx=210pt,bblly=410pt,bburx=630pt,bbury=550pt}
\\[5mm]
\rule{-13mm}{0mm} 
\rule{1mm}{0mm}(c) &\hspace*{-.1cm}\rule{-2mm}{0mm}(d) 
\end{tabular}}
\]
\end{minipage}
\vspace*{1mm}

\noindent
\begin{minipage}{16.cm}
\[
\mbox{
\hspace*{1cm}
\begin{tabular}{cc}
\psfig{figure=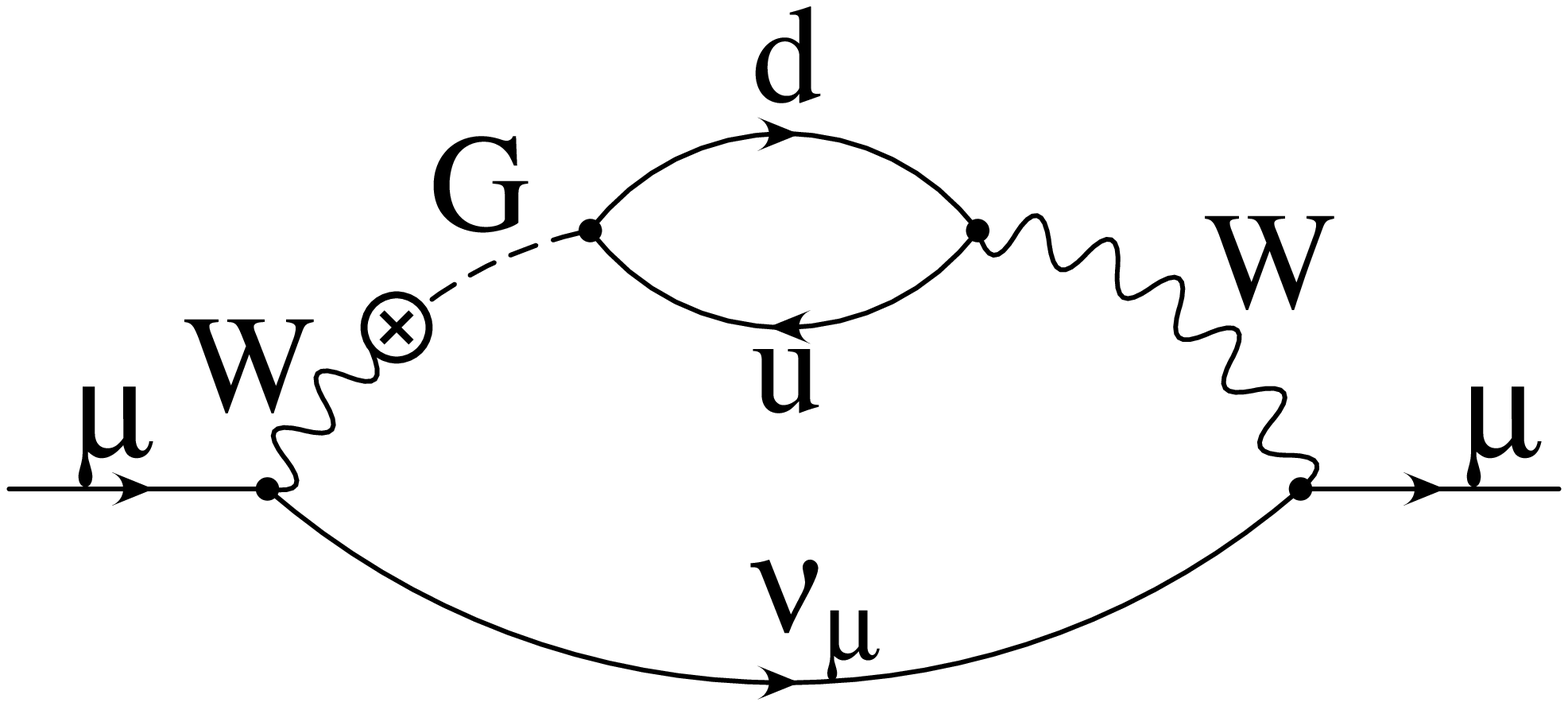,width=20mm,bbllx=210pt,bblly=410pt,bburx=630pt,bbury=550pt}
&\hspace*{.6cm}
\psfig{figure=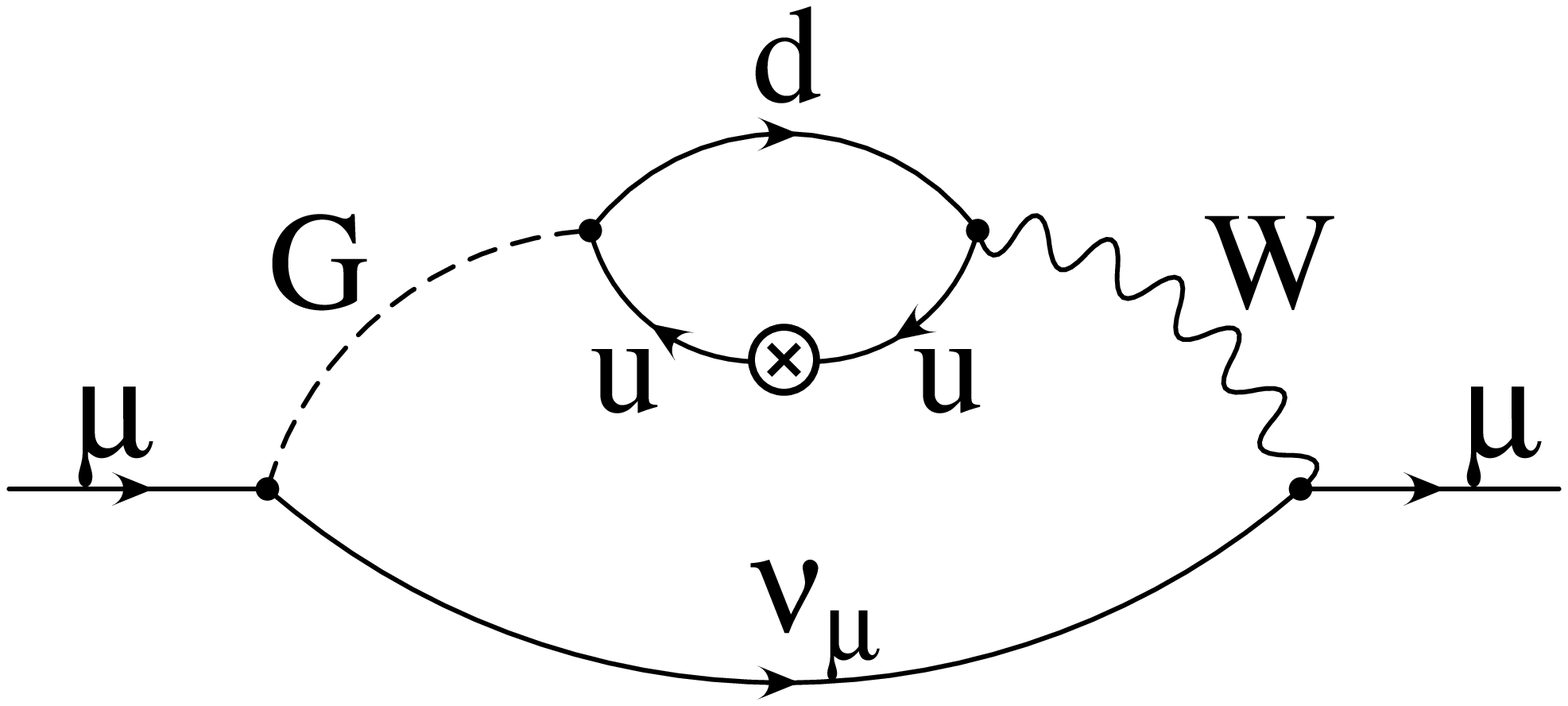,width=20mm,bbllx=210pt,bblly=410pt,bburx=630pt,bbury=550pt}
\\[5mm]
\rule{-13mm}{0mm} 
\rule{1mm}{0mm}(e) &\hspace*{-.1cm}\rule{-2mm}{0mm}(f) 
\end{tabular}}
\]
\end{minipage}
\vspace*{1mm}

\noindent
\begin{minipage}{16.cm}
\[
\mbox{
\hspace*{1cm}
\begin{tabular}{cc}
\psfig{figure=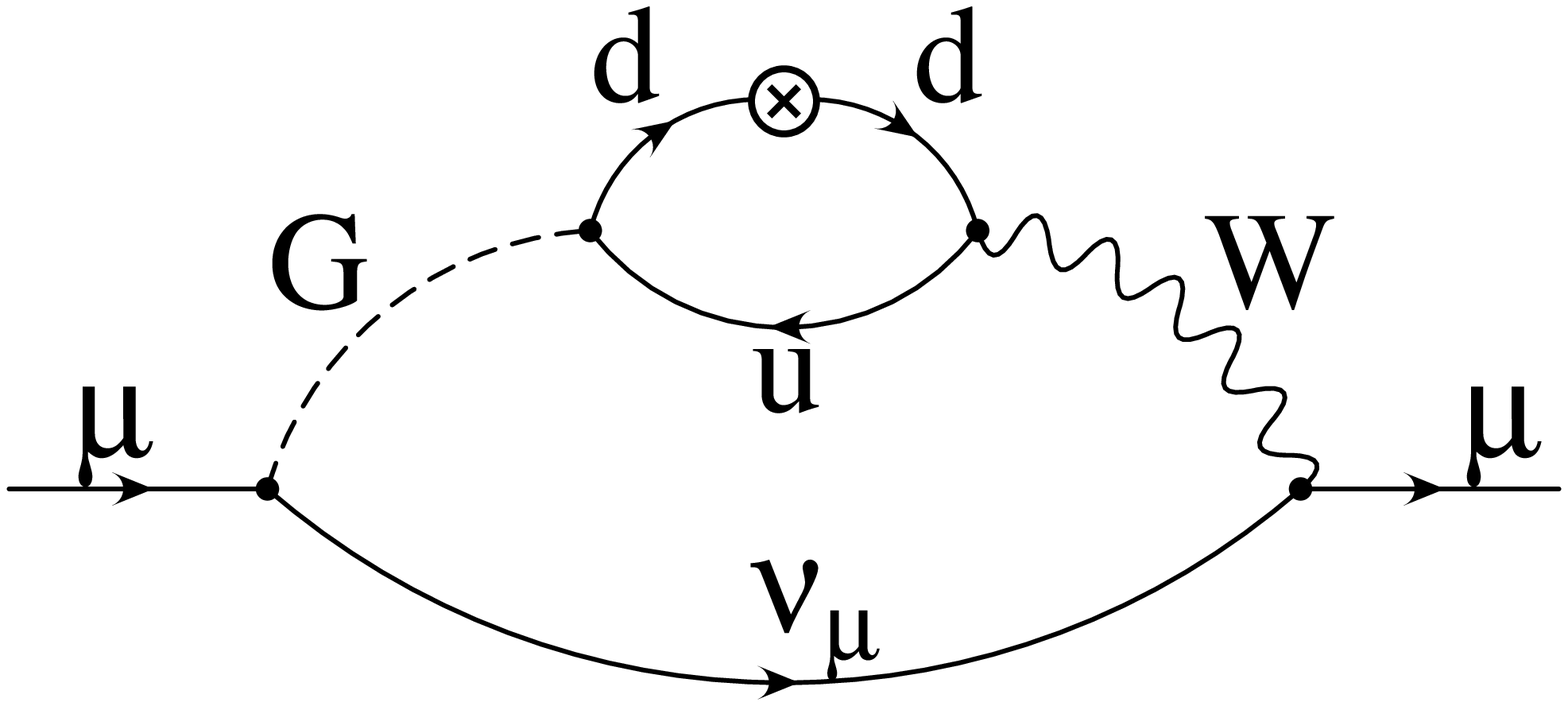,width=20mm,bbllx=210pt,bblly=410pt,bburx=630pt,bbury=550pt}
&\hspace*{.6cm}
\psfig{figure=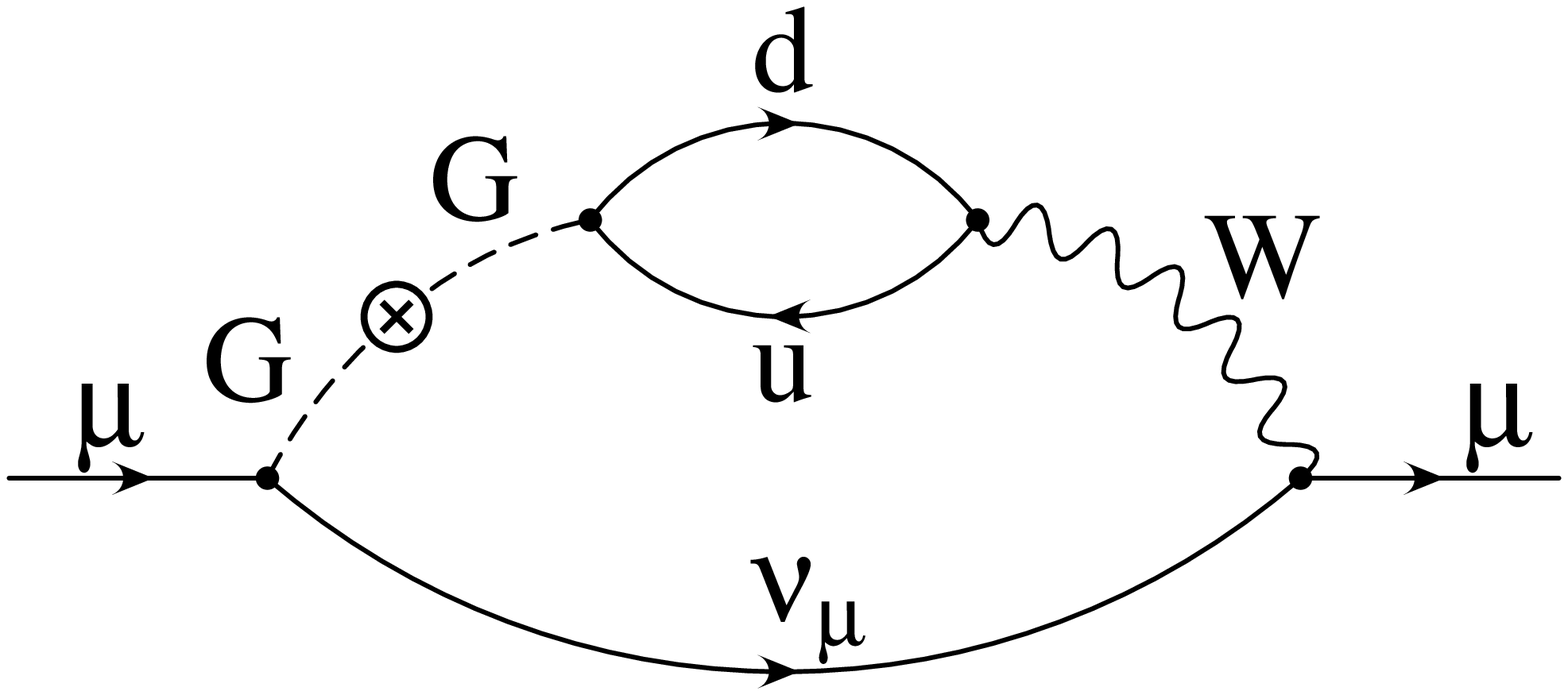,width=20mm,bbllx=210pt,bblly=410pt,bburx=630pt,bbury=550pt}
\\[5mm]
\rule{-13mm}{0mm} 
\rule{1mm}{0mm}(g) &\hspace*{-.1cm}\rule{-2mm}{0mm}(h) 
\end{tabular}}
\]
\end{minipage}
\vspace*{1mm}

\noindent
\begin{minipage}{16.cm}
\[
\mbox{
\hspace*{1cm}
\begin{tabular}{cc}
\psfig{figure=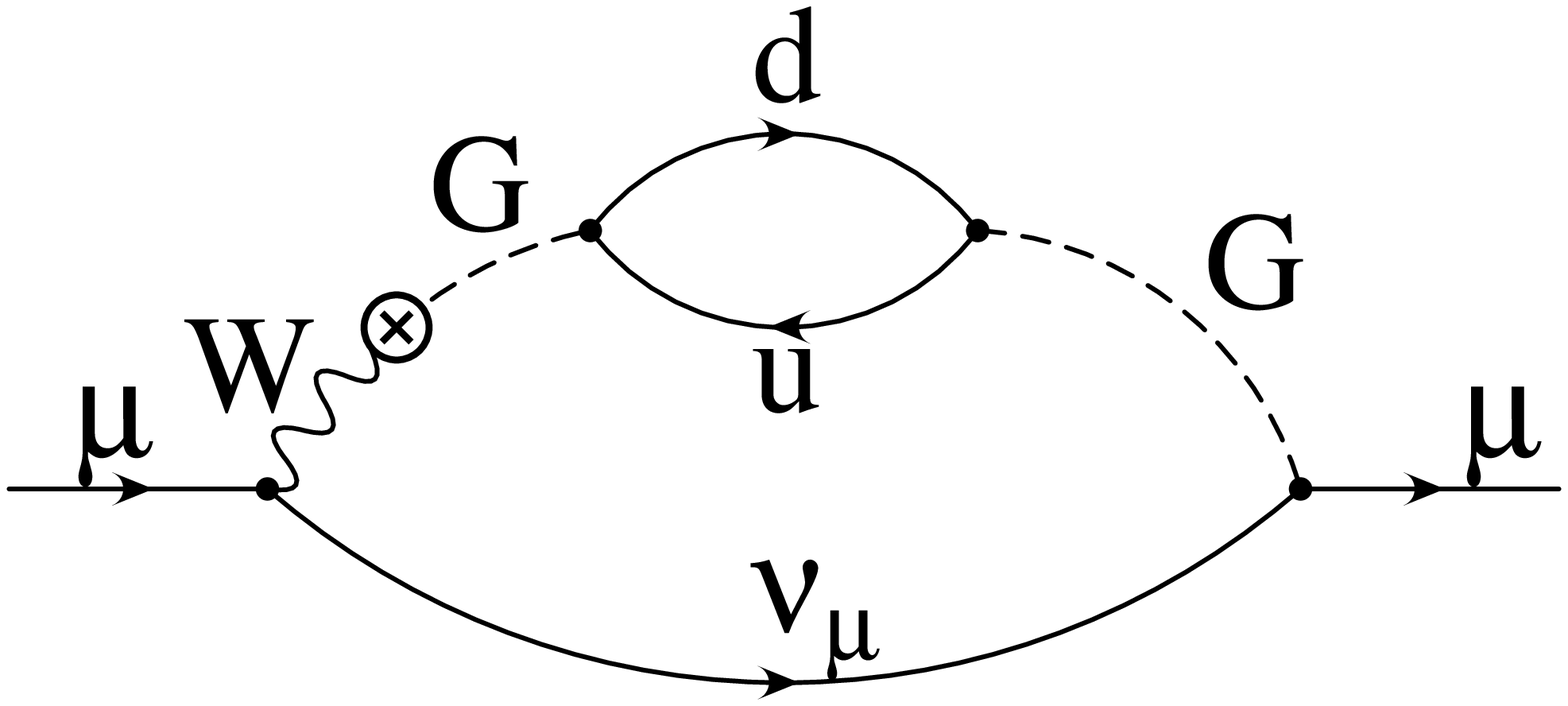,width=20mm,bbllx=210pt,bblly=410pt,bburx=630pt,bbury=550pt}
&\hspace*{.6cm}
\psfig{figure=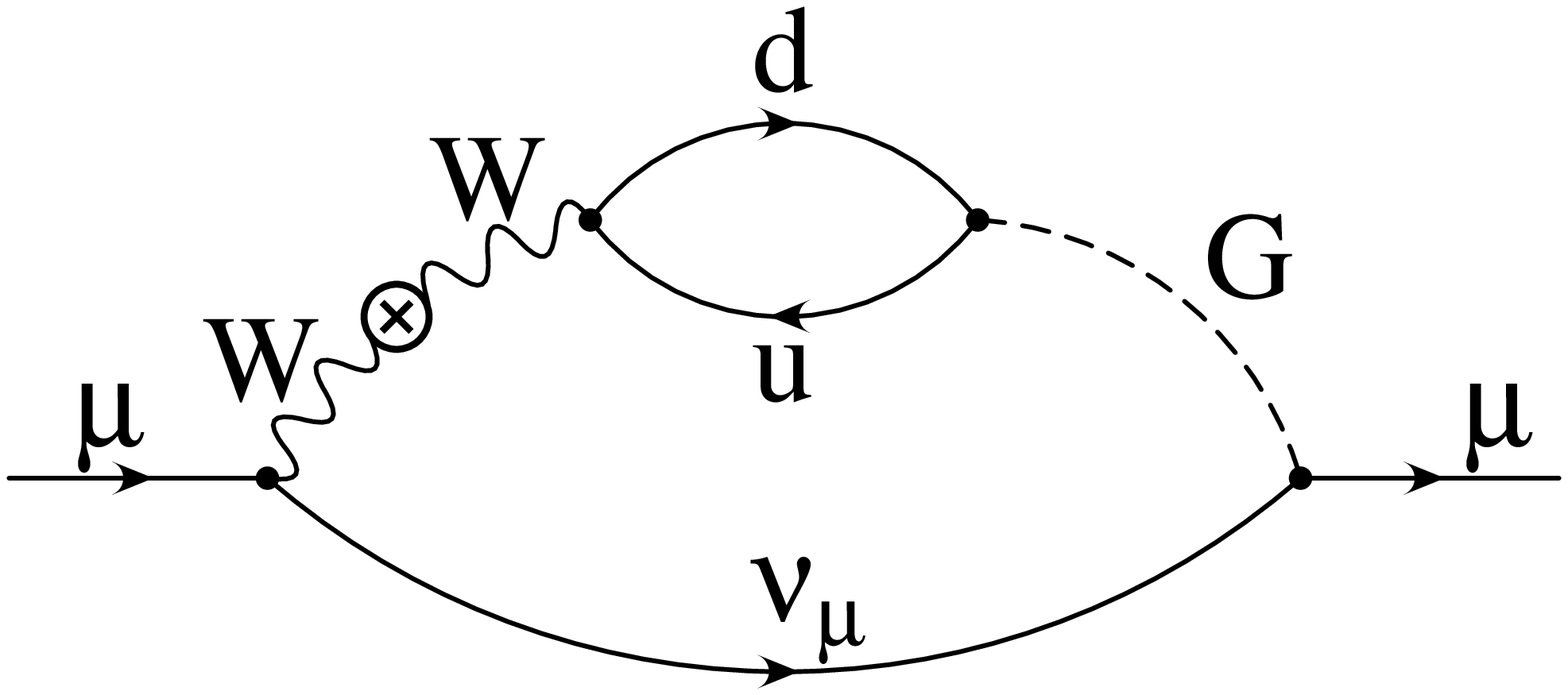,width=20mm,bbllx=210pt,bblly=410pt,bburx=630pt,bbury=550pt}
\\[5mm]
\rule{-13mm}{0mm} 
\rule{1mm}{0mm}(i) &\hspace*{-.1cm}\rule{-2mm}{0mm}(j) 
\end{tabular}}
\]
\end{minipage}

\vspace{.3cm}
\noindent 
Figure 3: Diagrams with a charged boson and a fermion loop 
contributing to the muon g-2.
Crossed circles denote interactions with an external photon.

\subsection*{Projector for g-2 form factor}

\noindent
An efficient projector for g-2 calculations has been obtained in
ref.\cite{ros90}. Here we extend their formulae
to $d=4-2 \ep$ dimensions.

\vspace{.9cm}
\begin{minipage}{10.cm}
\hspace{1.8cm}
\psfig{figure=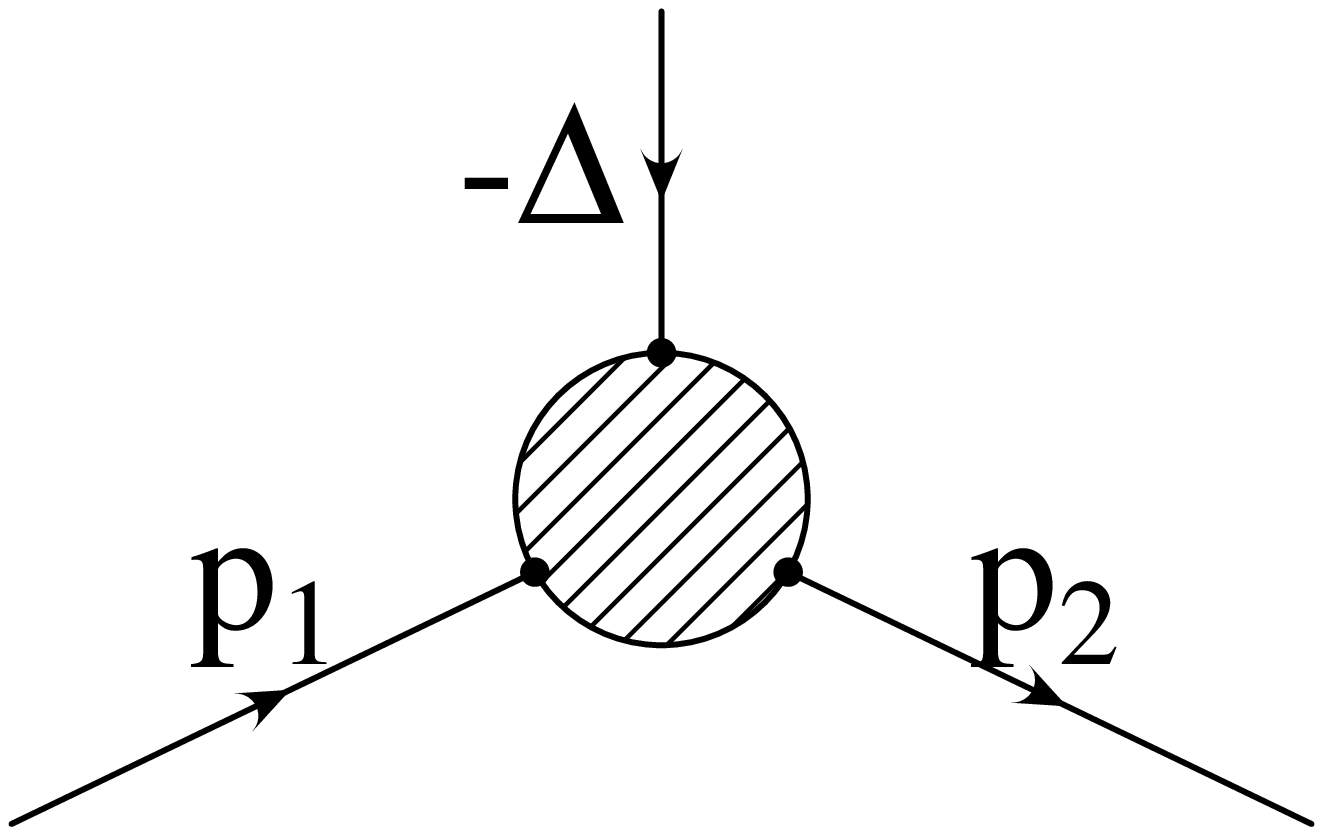,width=30mm,bbllx=210pt,bblly=410pt,bburx=630pt,bbury=550pt}
\end{minipage}
\vspace{.3cm}

Figure 4: Definition of momentum flow 
\vspace{.3cm} \\
We consider the most general matrix element of a
current between spin 1/2 fermions
\be
\langle\alpha_f|J_\mu(x)|\alpha_i\rangle &=& \\
&&\hspace{-3.0cm}    {\bar u}_f(p_2)\left[F_1(t)
\gamma_\mu  
-{i\over 2m}F_2(t)\sigma_{\mu\nu}\Delta^\nu
+{1\over m}F_3(t) \Delta_\mu  \right. 
 \nonumber\\
&&\hspace{-3.1cm} \left.+ \gamma_5\left(G_1(t) \gamma_\mu
-{i\over 2m}G_2(t)\sigma_{\mu\nu}\Delta^\nu
+{1\over m}G_3(t) \Delta_\mu \right)\right]\nonumber \\
&&\hspace{-3.1cm}\cdot u_i(p_1) e^{i\Delta x} \nonumber
\ee
with $\Delta = p_1-p_2$ and $\Delta^2 = t$. For on-shell external fermions we have 
\be
p_1^2 = p_2^2 = m^2
\ee
Introducing $p = {1\over2}(p_1+p_2)$ we obtain in addition:
\be
p^2 = {1\over4}(4 m^2-t)\:, \hspace{1cm} p\cdot\Delta = 0 \, .
\ee

Conservation of the electromagnetic current requires $F_3(t) = 0$. 
$F_1(t)$ is the charge
form factor, $F_2(t)$ the magnetic moment form factor, $G_2(t)$ the 
electric moment form factor.
The anomalous magnetic moment of the fermion $a$ is given by
\be
a \equiv {1\over2}(g-2) = F_2(0)
\ee

In order to extract the magnetic moment form factor, one can
introduce a projection operator
\be
N_\mu = \\
&&\hspace{-1.5cm}(\psla_1+m)\left[g_1\gamma_\mu-{1\over m}g_2 p_\mu
        -{1\over m}g_3 \Delta_\mu \right](\psla_2+m) \nonumber
\ee

In order to determine the coefficients $g_i$ we take the trace of
$N_\mu M^\mu$:
\be
\lefteqn{{\rm Tr}(N_\mu M^\mu) = 
   \left\{\left[8 m^2+2t(d-2)\right]g_1  \right.\nonumber} \\
&& \left.\hspace{1.7cm}+ 2(t-4m^2)g_2\right\}F_1(t) \nonumber\\
&& \hspace{-.6cm}+\left[2t(d-1)g_1+ t{(t-4 m^2)\over 2m^2}g_2\right]F_2(t) \nonumber\\
&& \hspace{-.6cm} +\left[2t{(t-4 m^2)\over m^2}g_3\right] F_3(t)
\ee

This leads us to the following system of equations for $F_2(t)$:
\be
0 &=& \left[8 m^2+2t(d-2)\right]g_1+ 2(t-4m^2)g_2 \nonumber\\
1 &=& 2t(d-1)g_1+ t{(t-4 m^2)\over 2m^2}g_2
\ee
and we can put $g_3=0$ already at this step.
The solution is
\be
g_1 &=& {-2m^2\over (d-2)\, t (t-4m^2)} \\
g_2 &=& {2m^2 \left[4 m^2+(d-2)t\right]\over (d-2)\, t (t-4m^2)^2 }
\ee
and we find
\be
F_2(t) &=& {-2m^2\over (d-2)\, t (t-4m^2)} {\rm Tr}\left[
  (\psla_1+m)\right. \times\nonumber \\
&&\hspace{-1.3cm}\left.\left(\gamma_\mu
  + {4 m^2+(d-2)t\over m(t-4m^2) } p_\mu\right)
         (\psla_2+m) M^\mu \right] 
\ee

It is also possible to extract directly the anomaly $F_2(0)$. As a
first step the general amplitude $M_\mu$ can then be expanded to first
order in $\Delta_\mu$:
\be
M_\mu(p,\Delta) &\approx& M_\mu(p,0)
+\left. \Delta_\nu{\partial\over\partial\Delta_\nu}
M_\mu(p,\Delta)\right|_{\Delta=0} \nonumber\\
&& \equiv V_\mu(p)+\Delta_\nu T^\nu_\mu(p) \, .
\ee

The next step is to average over the spatial directions of $\Delta$
with the formulas
\be
\langle\Delta_\mu\Delta_\nu\rangle &=& 
{1\over d-1}\Delta^2\left(g_{\mu\nu}-{p_\mu p_\nu\over p^2}\right)\,
,\nonumber\\
\langle\Delta_\mu\rangle &=& 0 \; ; 
\ee
after that, the limit $t\to 0$ can be taken. The result is
\be
\lefteqn{a ={1\over 2(d-1)(d-2)m^2} \times }\\
&&{\rm Tr}\left\{{d-2\over2}
 \left[m^2\gamma_\mu - d p_\mu \psla - (d-1)mp_\mu\right]V^\mu
\right.\nonumber\\
&&\left.\hspace{0.4cm}
+{m\over4}(\psla+m)[\gamma_\nu,\gamma_\mu](\psla+m)T^{\mu\nu}
\right\} \nonumber
\ee

\section*{Acknowledgement}
We thank Professor W. Marciano for collaboration on the topic of this
talk. 
This research was supported by BMBF 057KA92P and by ``Graduiertenkolleg
Elementarteilchenphysik'' at the University of Karlsruhe.

\end{document}